\pdfoutput=1
\RequirePackage{amsmath}
\documentclass[runningheads]{llncs}
\usepackage{cite}
\usepackage{graphicx}
\usepackage{color}
\usepackage{algorithm,algorithmic}
\usepackage[caption=false]{subfig} 
\usepackage[hyphens]{url}
\usepackage{booktabs}
\usepackage{multirow}
\usepackage{array}
\usepackage{tabularx} 
\usepackage[misc,geometry]{ifsym} 

\usepackage[colorlinks=true,allcolors=blue]{hyperref}

\begin{document}
\title{A Lightweight Music Texture Transfer System}
%
%
\author{Xutan Peng\inst{1,2}\textsuperscript{(\Letter)} \and
Chen Li\inst{1,2} \and
Zhi Cai\inst{1,2}\and
Faqiang Shi\inst{1,3}
Yidan Liu\inst{44}\and 
Jianxin Li\inst{1,2}
}
\authorrunning{X. Peng et al.}
%
\institute{Beijing Advanced Innovation Center for Big Data and Brain Computing
\email{pengxt@act.buaa.edu.cn} \and
SKLSDE Lab, Beihang University \and 
State Key Laboratory of VR Technology and Systems, Beihang University \and 
Department of Psychology, Beihang University }

\maketitle              
\setcounter{footnote}{0}
\begin{abstract}
Deep learning researches on the transformation problems for image and text have raised great attention. However, present methods for music feature transfer using neural networks are far from practical application. In this paper, we initiatively propose a novel system for transferring the texture of music, and release it as an open source project. Its core algorithm is composed of a converter which represents sounds as texture spectra, a corresponding reconstructor and a feed-forward transfer network. We evaluate this system from multiple perspectives, and experimental results reveal that it achieves convincing results in both sound effects and computational performance.

\end{abstract}
\section{Introduction}
Currently, great amounts of work has verified the power of Deep Neural Networks (DNNs) applicated in multimedia area.
Among them, as a popular artificial intelligence task, transformation of input data has obtained some competitive results.
Particularly, some specific neural networks for artistic style transfer can novelly generate a high-quality image through combining the content with the style information from two different inputs\cite{Gatys,Johnson2016Perceptual,CycleGAN2017}.

Through utilizing corresponding representation methods and modifying their convolutional neural structures, many algorithms in other fields (e.g., text) have also achieved competitive effects on transferring various features like style\cite{nlptf,p_style} or sentiment\cite{sentiment}.

However, different from endeavor in transferring image or text, the transformation of music features, exclusively restrained by the field per se, is still in its infancy. More specially, as a sequential and continual signal, music is significantly different from image (non-time-series) or text (discrete), thus mature algorithms adopted in other fields cannot be used directly. Present methods neglect the problems mentioned above and only consider some specific factors as music features, such as frequency, channel, etc. The outputs of these further attempts of transformation based on source's statistic features are far from satisfaction.

In the field of music, `style' transfer has been widely investigated, but it's still poorly defined. Therefore, in this paper, instead of `style', we regard \emph{texture} as our transfer object, i.e., the collective temporal homogeneity of acoustic events\cite{goldstein2014sensation}. In musical post-production, the transformation and modification of texture have long been a common but time-consuming process. About this, Yu et al. found specific repeated patterns in short-time Fourier spectrograms of solo phrases of instruments, reflecting the texture of music\cite{texture_images}. 
Thereupon, our motivation is to firstly develop a novel reconstructive spectral representation on time-frequency for audio signal like music. It can not only preserve content information but also distinguish texture features. Secondly, to further achieve successful texture transfer, we selectively exploit the convolution structure that has succeeded in other fields, and make adaptations for the integration of our end-to-end model. Lastly, to generate music from the processed spectral representation, we design a reconstruction algorithm for final music output.

\begin{figure*}[h]
\begin{center}
 \centerline{\includegraphics[width=\textwidth]{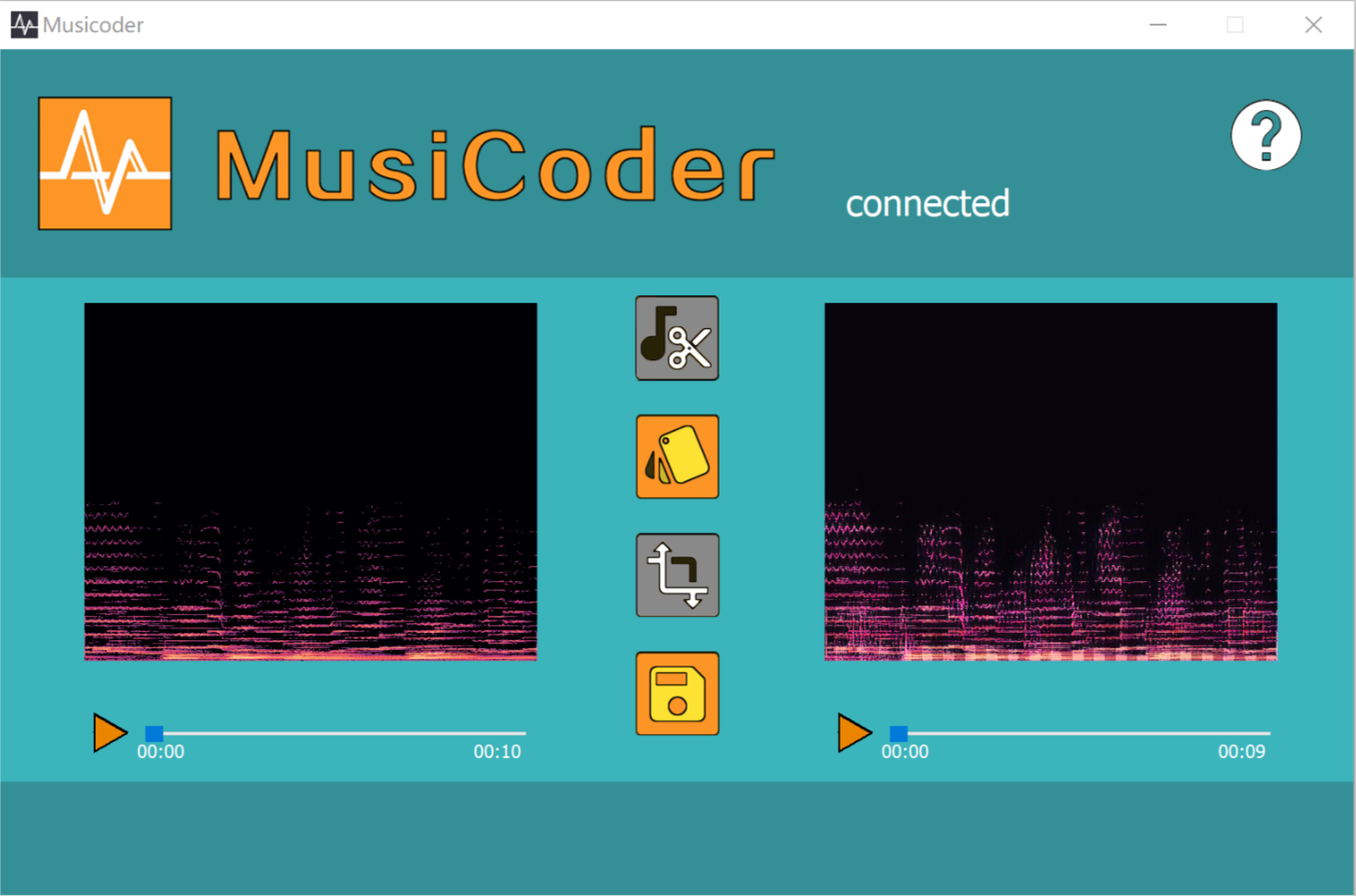}}
 \caption{The user interface of MusiCoder's PC client. This client provides entry to our online texture transfer service. The left window is for interactively music-inputing while both windows display spectral images of input and output music. For each transfer task, this client allows users to select and preview a 10-second clip of original samples. Users are able to opt for target texture as well as final quality before each run. Transfered music can be easily saved to local path.}
 \label{fig:musicoder_ui}
 \end{center}
\end{figure*}

By applying the proposed network for texture transformation of music samples, we validate that our method has compelling application value. To further assess this model, a demo termed \emph{MusiCoder} is  deployed and evaluated. The user interface of its PC client is shown in Fig.~\ref{fig:musicoder_ui}. For reproducibility, we release our code as an open source project on GitHub\footnote{\url{https://github.com/Pzoom522/MusiCoder}}.

To sum up, the main contributions of our work are listed as follows:

\let\labelitemi\labelitemii
\begin{itemize}
\item With integration of our novel reconstructive spectral representation and transformation network, we propose an end-to-end texture transfer algorithm;
\item To the best of our knowledge, we first develop and deploy a practical music texture transfer system. It can also be utilized for texture synthesis (Sect.~\ref{ts});
\item We propose novel metrics to evaluate transformation of music features, which comprehensively assess the output quality and computational performance.
\end{itemize}

\section{Related Work}
The principles and approaches related to our model have been discussed in several pioneering studies. 

\subsubsection{Transformation for Image and Text.}
As the superset of texture transfer problems, a wide variety of models about transformation tasks have been sparked. In computer version, based on features extracted from pre-trained Convolutional Neural Networks (CNNs), Gatys et al. perform artistic style transfer on image by jointly minimizing the loss of content and style\cite{Gatys}.
However, its high computational expense is a burden. Johnson et al.\cite{Johnson2016Perceptual} demonstrate a feed-forward network to provide approximate solutions for similar optimization problem almost in real time. The latest introduction of circularity has inspired many popular constraints for more universal feature transfer such as CycleGAN\cite{CycleGAN2017} and StarGAN\cite{stargan}.

In natural language processing, transformation of features (e.g., style and sentiment) is regarded as controlled text generation tasks. Recent work includes stylization on parallel data\cite{p_style} and unsupervised sequence-to-sequence transfer using non-parallel data\cite{nlptf,sentiment}. Their best results are now highly correlate to human judgments.

\subsubsection{Transformation for Audio and Music.}
Scarce breakthrough of feature transfer has been made in audio or music. Inspired by research progress in image style transfer, some approaches discuss the music `style' transfer which is redefined as cover generation\cite{Malik,Mital}. They directly adopt modified image transfer algorithms to obtain audio or music `style' transfer results. Despite that the output music piece changes its `style' to some extent, the overall transferring effect remains unsatisfactory. 

Other approaches take different tacks to perform music feature transfer. Wyse performs texture synthesis and `style' transfer by utilizing a single-layer random-weighted network with 4096 different convolutional kernels after investigating the formal donation between spectrum and image\cite{Wyse2017}. Barry et al. adopt similar idea by demonstrating Mel and Constant-Q Transform apart from the original Short Term Fourier Transform (STFT)\cite{barry2018style}. These methods, however, fail to clearly discriminate content and `style', and have poor computational performance.

\subsubsection{Generative Music Models.} 
Recent advances in generative music models include WaveNet\cite{Oord2016WaveNetAG} and DeepBach\cite{deepbach}. These more complicated models offer new possibilities for music transformation. Specially, a very recent model based on WaveNet\cite{Oord2016WaveNetAG} architecture is proposed by Mor et al.\cite{fair}, and it impressively produces high-quality results. Successful as it is, this model targets to problems of higher level, which clearly distances itself from texture transfer methods. Meanwhile, it has a drawback of limited feasibility to build real-world applications, owing to the structural and subsequently computational complexity of present approaches for generating music.

To the best of our knowledge, our model is the first practical texture transfer method for music, making it ahead of other approaches in both efficiency and performance. 

\section{Methodology}

\subsection{Problem Definition and Notations}
Given a music and audio pair $(M_i,A_i)$, we have $r_t(M_i,A_i)=True$ iff. $M_i$ and $A_i$ share common recognizable texture. Similarly, $r_c(M_j,N_j)=True$ exists iff. $M_j$ and another piece of music $N_j$ are regarded as different versions of the same music content. Given a pair of music and audio $(M_c,A_t)$, music texture transfer is to generate music piece $M_t$  which satisfies $r_t(M_t,A_t) \land r_c(M_t,M_c)=True$.

\begin{figure*}[ht]
\begin{center}
 \includegraphics[width=\textwidth,trim=3 2 2 2,clip]{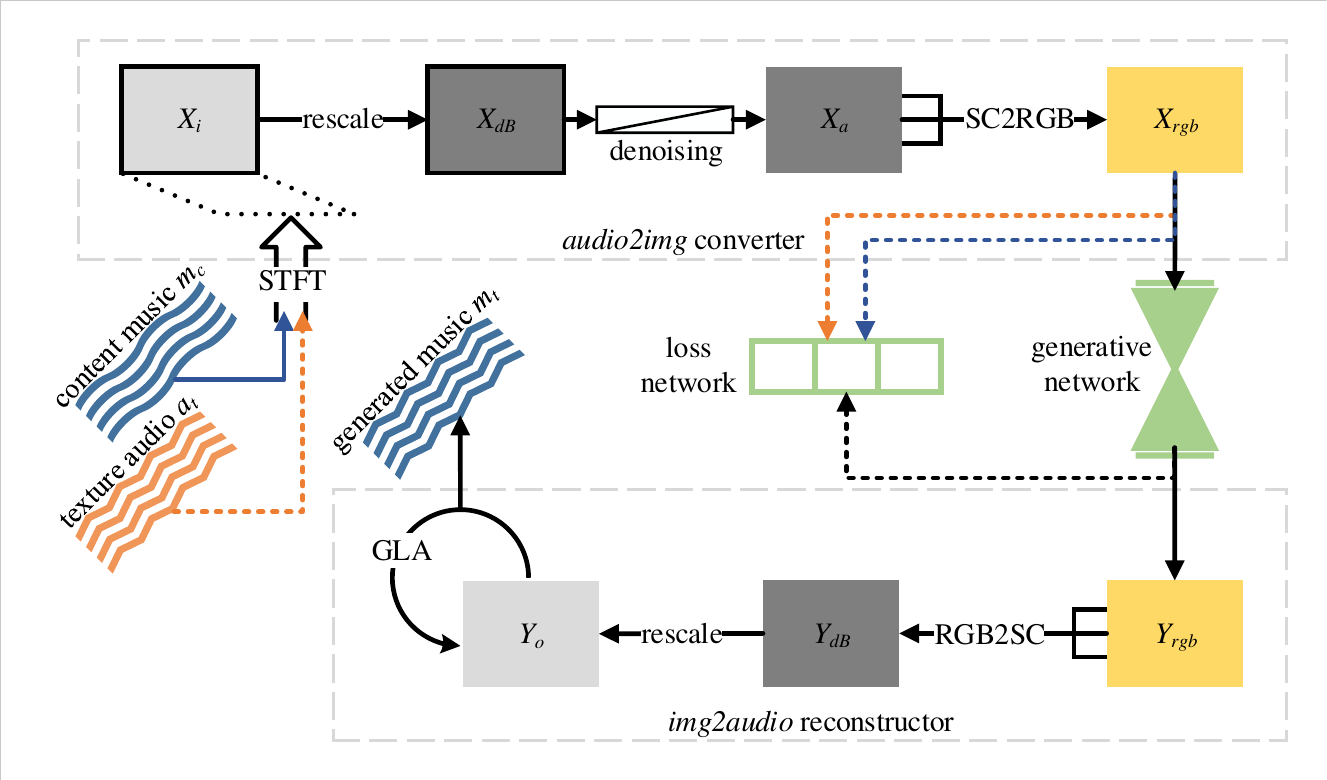}
 \caption{The overview architecture of our core algorithm. Loss network is utilized during training and is not required during feed-forward process (production environment). The dashed lines indicate the data flow that only appears in training.}
 \label{fig:overview}
 \end{center}
\end{figure*}

\subsection{Overall Architecture and Components}
The overall architecture of our core texture transfer algorithm is illustrated in Fig.~\ref{fig:overview}. For each run of texture transfer, we first input $M_c$ to the $audio2img$ converter which returns corresponding spectral representation. Then this spectrum is fed into a pre-trained feed-forward generative network. Lastly, the $img2audio$\ reconstructor restores generated spectrum to $M_t$. The detailed structure of each component is presented in the following subsections.

\subsubsection{$audio2img$ Converter.}\label{$audio2img$}

\begin{figure*}[ht]
\begin{center}
\subfloat[$X_{i}$\label{fig:xi}]{\includegraphics[width=0.24\textwidth, height=2.4cm]{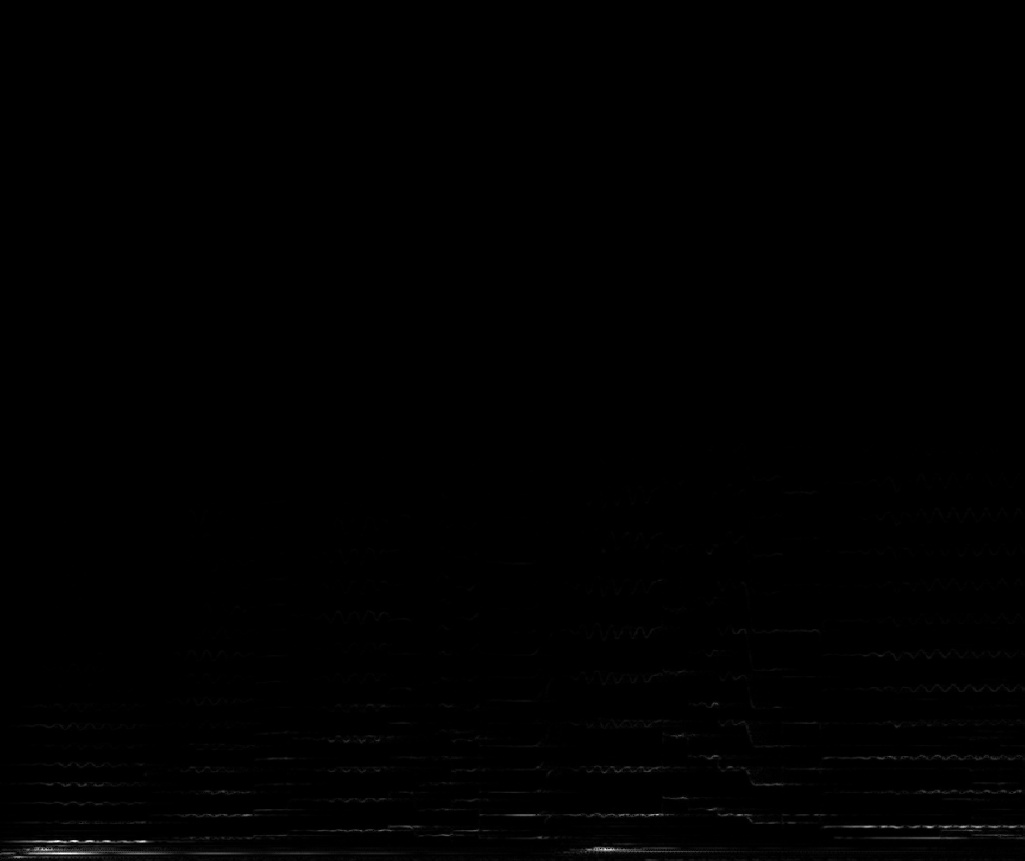}} \
\subfloat[$X_{dB}$\label{fig:xdb}]{\includegraphics[width=0.24\textwidth, height=2.4cm]{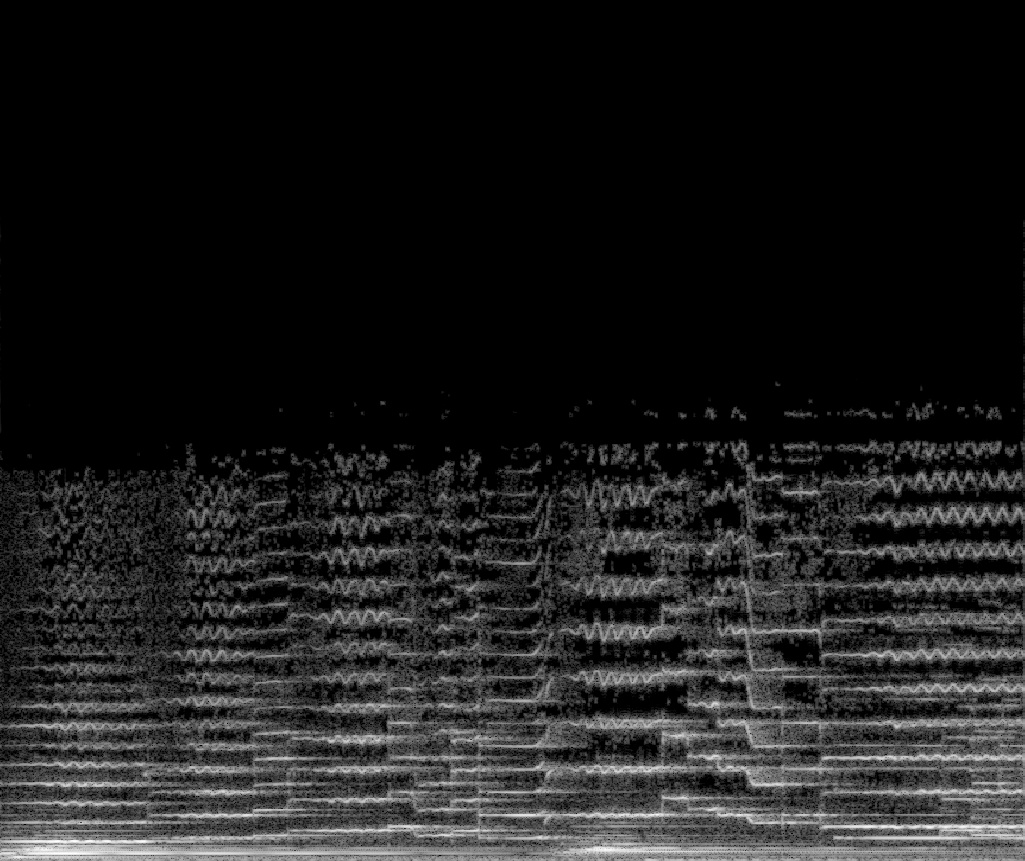}} \ 
\subfloat[$X_{a}$\label{fig:xa}]{\includegraphics[width=0.24\textwidth, height=2.4cm]{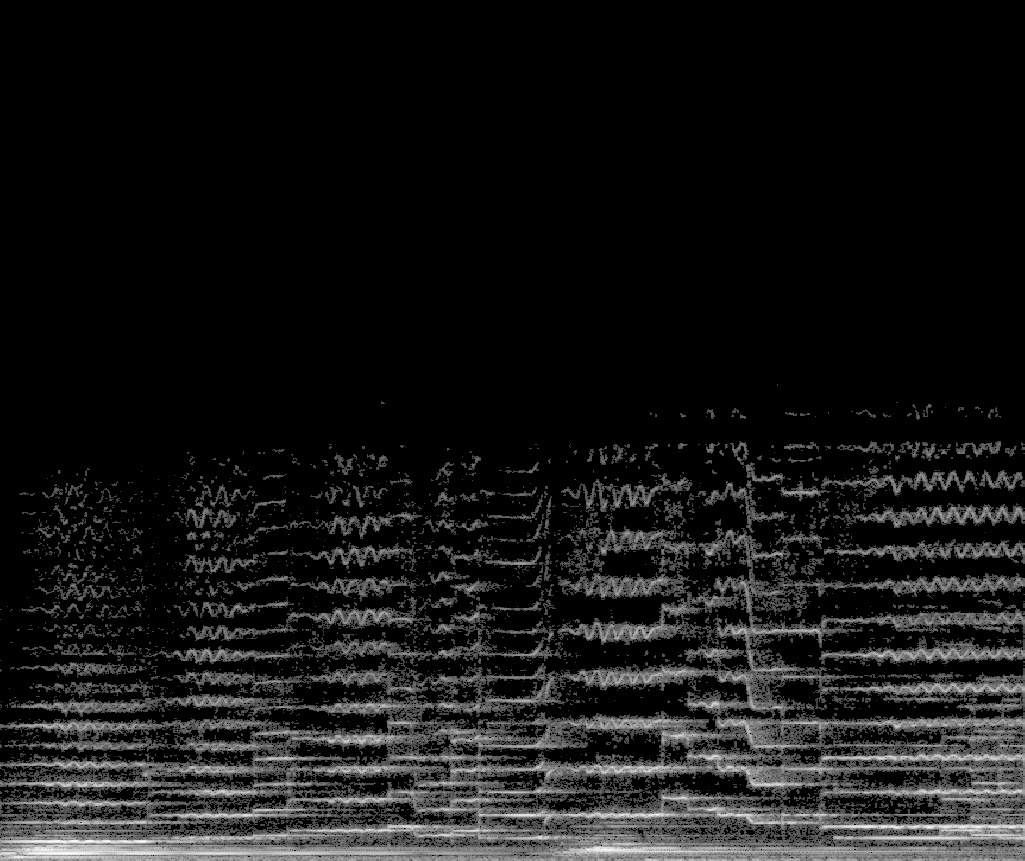}} \ 
\subfloat[$X_{dB}-X_{a}$\label{fig:lost}]{\includegraphics[width=0.24\textwidth, height=2.4cm]{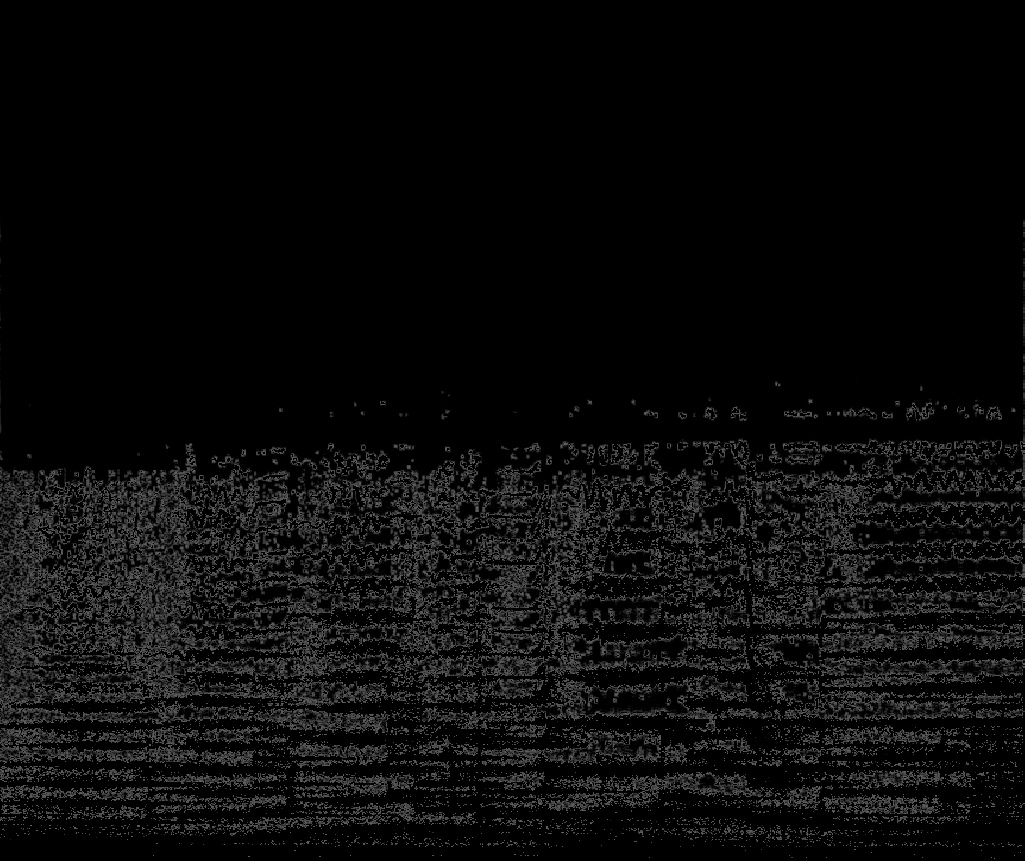}} \\ 
\subfloat[$X_{rgb}$\label{fig:xrgb}]{\includegraphics[width=0.24\textwidth, height=2.4cm]{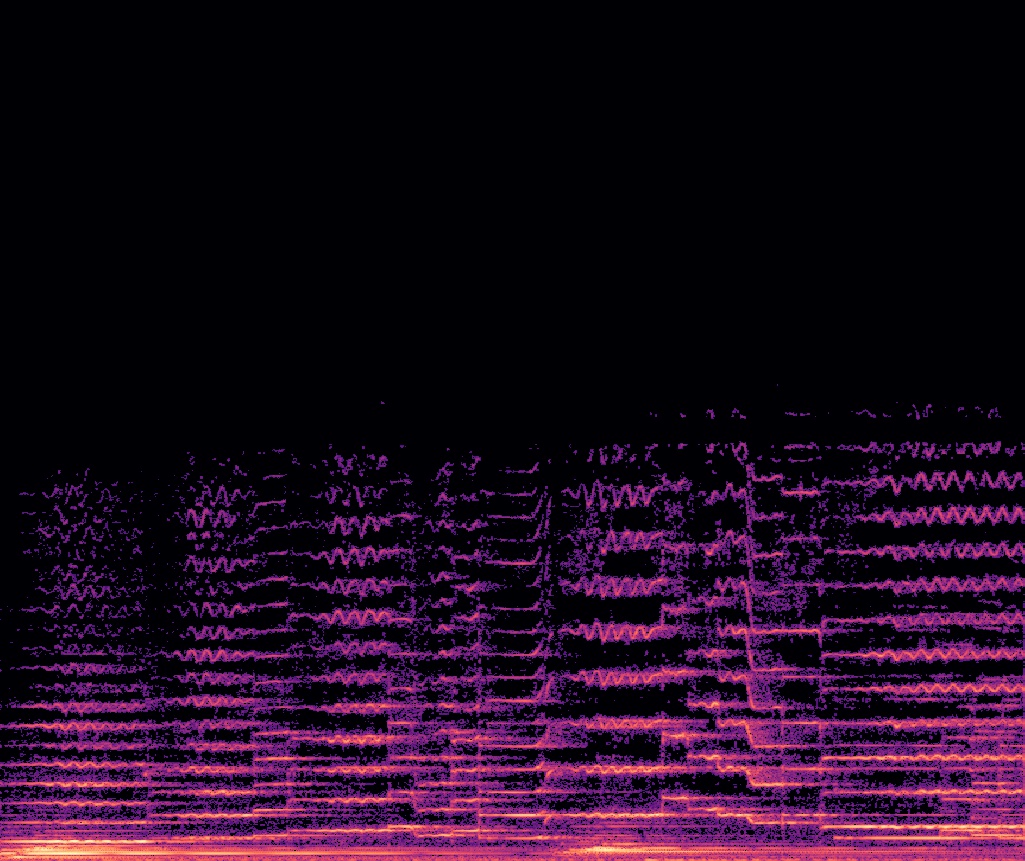}} \ 
\subfloat[$Y_{rgb}$\label{fig:yrgb}]{\includegraphics[width=0.24\textwidth, height=2.4cm]{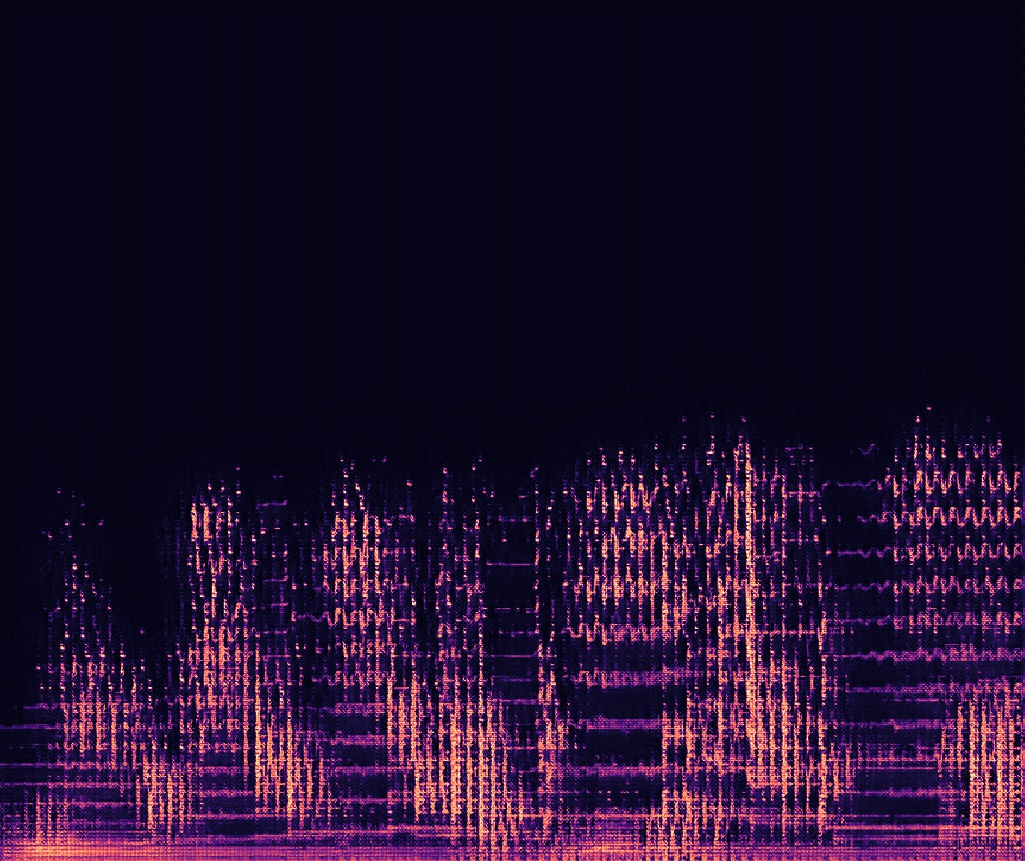}} \ 
\subfloat[$Y_{dB}$\label{fig:ydb}]{\includegraphics[width=0.24\textwidth, height=2.4cm]{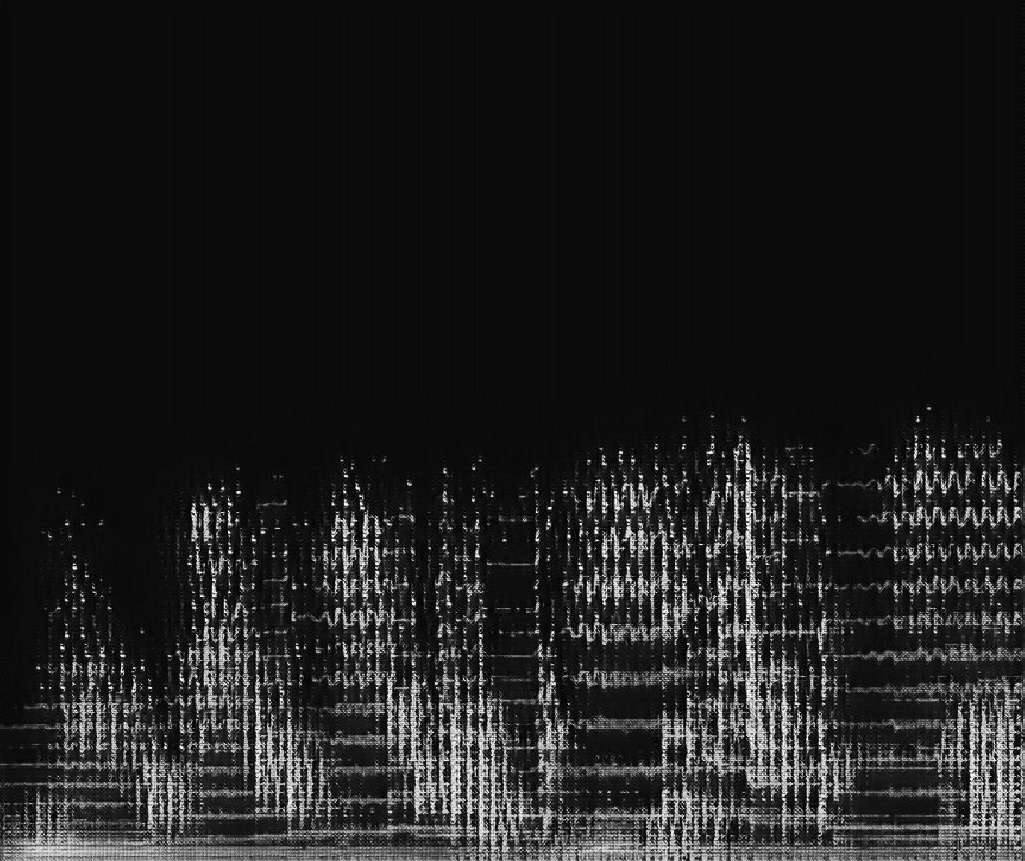}} \ 
\subfloat[$Y_o$\label{fig:yo}]{\includegraphics[width=0.24\textwidth, height=2.4cm]{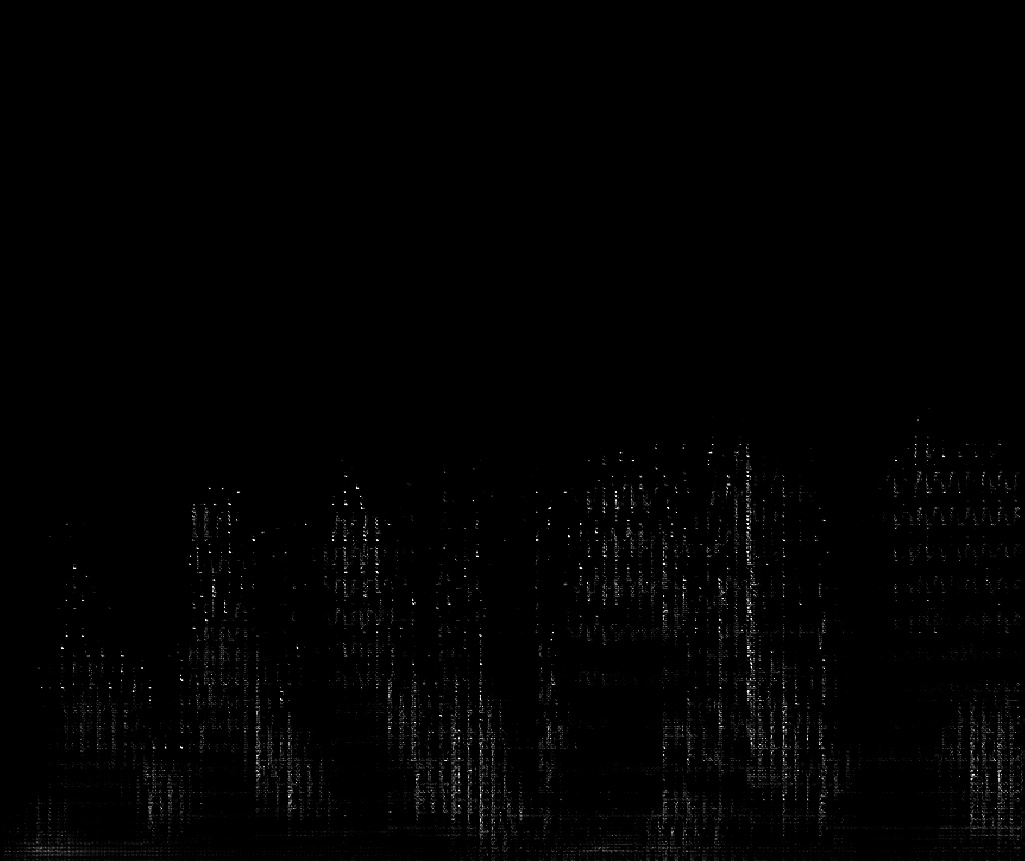}} \\ 
  \caption{Spectra of different intermediates using a 10-second sample during the feed-forward texture transfer process. See Sect.~\ref{experiment_setup} for the detailed network configuration and training parameters. Here, we select `Water' (Fig.\ref{fig:water}) as target texture. (a)(b)(c)(e) exhibit the intermediates in $audio2img$ converter. (d) visualizes the loss introduced by denoising threshold, most of which doesn't contain non-negligible information in capturing features and restoring signal. (f)(g)(h) are intermediates produced in music reconstruction. Vertical lines in (f)(g) distinctly illustrate the texture characteristic of target texture.}
 \label{fig:spectra}  
 \end{center}
\end{figure*}

Given an acoustical piece $A_{i}$, by practicing the Fourier transform on successive frames, we can donate its phase and magnitude on time-frequency as:

\begin{equation}\label{stft}
S(m, \omega) = \sum_n x(n) w(n-m) e^{-j \omega n}
\end{equation}\\where $w(\cdot)$ is the Gaussian window centered around zero, and $x(\cdot)$ refers to the signal of $A_{i}$. We take the magnitude component as $X_{i}$. 

As shown in Fig.~\ref{fig:xi}, the spectrum plotting $X_{i}$ could reveal little information. Linearly growing amplitude is not in the full sense perceptually relevant to humans\cite{MCDERMOTT2011926}, therefore, based on Decibel, we rescale the spectrum into $X_{dB}$ as:

\begin{equation}\label{db}
X_{dB}=20 log(X_{i} / r)
\end{equation}\\where $r$ is the maximal value of $X_{i}$. See Fig.~\ref{fig:xdb} for the spectrum of $X_{dB}$.

The magnitude of rhythm information which tightly pertains to the reorganization of music content, is sharply larger than that of other information, e.g., ambient noise. Besides, in further implementation, it's noticed that the latter shows exerted detrimental effects on capturing texture and brings little improvement to audio signal reconstruction. As a result, we then design a heuristic denoising threshold mask which constructs spectrum $X_a$ as:

\begin{equation}\label{attentionA}
X_a=X_{dB} \odot H_{dB}+ \mathrm{min}(X_{dB})\lnot H_{dB}
\end{equation}

\begin{equation}\label{attentionB}
{H_{dB}}_{ij}=\begin{cases}
0,& {X_{dB}}_{ij}<\lambda \mathrm{min}(X_{dB})\\
1,& \text{otherwise}
\end{cases}
\end{equation}\\where $\lambda$ is a hyper-parameter which is in the interval of [0,1], $\odot$ donates the operation of Hadamard Product, and min$(\cdot)$ returns the minimal element in corresponding matrix. 

Unlike approaches which set up a channel for every single frequency\cite{barry2018style,Wyse2017}, in the succeeding transformation module, we map $X_{dB}$ into 3-channel $X_{{rgb}}$ so as to keep data in alignment.

\subsubsection{Feed-forward Generative Network.}\label{FGN}
To perform music texture transfer, a generative network which gets impressive results in dealing with image style transfer\cite{Johnson2016Perceptual} is employed as the basic architecture. 
In comparison to the original work, we utilize instance normalization\cite{Ulyanov2016InstanceNT} to achieve better results in task-completing. Our network consists of 3 layers of convolution and ReLU nonlinearities, 5 residual blocks, 3 transpose convolutional layers and a non-linear tanh layer which produces the output. Using activations at different layers of a pre-trained loss network, we calculate the content loss and texture loss between the generated output and our desired spectrum. They are donated as $L_{content}$ and $L_{texture}$ respectively:

\begin{equation}\label{Lcontent}
L_{content}=\frac{1}{2}\sum_{i,j}{(F_{ij}-P_{ij})}^2
\end{equation}

\begin{equation}\label{Ltexture}
L_{texture}=\frac{1}{2}\sum_{l=0}^L{(G_{ij}^l-A_{ij}^l)}^2
\end{equation}\\where $F_{ij}$ and $P_{ij}$ donate activations of the $i$th filter at position $j$ for the spectra of content and output respectively. $G_{ij}^l$ and $A_{ij}^l$ are layer $l$'s Gram Matrix of generated spectrum and texture spectrum, which is defined using feature map set $X$ as:
\begin{equation}\label{gram}
G_{ij}^l=\sum_{k}X_{ik}^lX_{jk}^l
\end{equation}

Let $L_{tv}$ donate the total variation regularizer which encourages spatial smoothness, then the full objective function of our transfer network is:
\begin{equation}\label{Ltotal}
L_{total}=\alpha L_{content}+ \beta L_{texture}+ \gamma L_{tv}
\end{equation}

During training, fixed spectrum of $M_t$ and spectra of a large-scale content music batch are fed into the network. We calculate the gradient via back-propagation of each training example and iteratively improve the network's weights to reduce the value of the loss function, making it possible for the trained generative network to apply certain texture to any given content spectrum with a single forward-propagation.

\subsubsection{$img2audio$\ Reconstructor.}
To reconstruct music with given spectrum $Y_{rgb}$, we have to firstly map it back from a 3-channel RGB matrix $M_{rgb}$ to single-channel $Y_{dB}$. In order to increase the processing speed, we design a conversion algorithm adopting the finite-difference method:

{\centering
\begin{minipage}{.79\linewidth}
  \begin{algorithm}[H]
 \caption{$RGB2SC$}
 \begin{algorithmic}
 \renewcommand{\algorithmicrequire}{\textbf{Input:}}
 \renewcommand{\algorithmicensure}{\textbf{Output:}}
 \REQUIRE The ascending 3-channel RGB list of selected color map, $C_m$; the 3-channel RGB spectrum, $M_{rgb}$
 \ENSURE  The single-channel spectrum, $M_{sc}$ 
 \\ \textit{Initialization} :
 \\ \textit{$C_{m-s} \gets \sum_{rgb} C_m$} 
 \\ \textit{$M_{rgb-s} \gets \sum_{rgb} M_{rgb}$}
 \\ \textit{$M_{sc}, M_{one} \gets \neg (M_{s-rgb}-M_{rgb-s})$}
 \\ 
  \FOR {$i = 0$ to $(len(C_m)-2)$}
  \STATE $d \gets C_{m-s}[i+1]-C_{m-s}[i]$
  \STATE $M_{rgb-s} \gets M_{rgb-s}-d \cdot M_{one}$
  \STATE $M_{sc}[M_{rgb-s}<0] \gets i/(len(C_m)-1)$
  \STATE $M_{rgb-s}[M_{rgb-s}<0] \gets 3$
  \ENDFOR
 \RETURN $M_{sc}$ 
 \end{algorithmic} 
 \end{algorithm}
\end{minipage}
\par
}
\bigskip
Then, after manipulating the approximate inverse operation of Decibel calculation, we scale $Y_{dB}$ back along its frequency axis:
\begin{equation}\label{idb}
Y_o = {\sqrt{10}}^{ Y_{dB} + \frac{log(r)}{10}}
\end{equation}\\where $r$ is same as that in $audio2img$ converter.

As for the recovery of phase information, we adopt Griffin-Lim Algorithm (GLA) which iteratively computes STFT and inverse-STFT until convergence\cite{GLA}. With adjusting the volume of our final output to the initial value, we produce the generated audio $A_o$.

\section{Experiment}\label{experiment}
To validate our proposed system, we deployed its production environment on a cloud server with economical volume\footnote{CPU: a mononuclear Intel\textsuperscript{\textregistered} Xeon\textsuperscript{\textregistered} E5-26xx v4 $||$ RAM: 4 GB}. We did experiments to generate music that integrated the content of input music with the texture of given audio, as well as to assess our system in both output quality and computational expense. Our experimental examples are freely accessible\footnote{\url{https://pzoom522.github.io/MusiCoder/audition/}}.

\subsection{Experimental Setup}\label{experiment_setup}
We trained our network on the Free Music Archive (FMA) dataset\cite{fma_dataset}. We trisected 106,574 tracks (30 seconds each), and loaded them with their native sampling rates. We set the FFT window size to be 2048 and $\lambda$ of denoising threshold to be 0.618. Audio signal was converted into 1025$\times$862 images using our $audio2img$ converter. For training the feed-forward generative network, we utilized a batch size of 16 for 10 epochs over the training data. The learning rate we set was $0.001$. To compute loss function, we adopted a pre-trained VGG-19\cite{VGG19} as our loss network. We set $7.5$, $500$ and $200$ as the weights of $L_{content}$, $L_{texture}$ and $L_{tv}$ respectively for texture transfer. As for $img2audio$ reconstructor, the number of iteration in GLA was 100.

\subsection{Datasets}\label{datasets}
For texture audio, we selected $S_{texture} = \{\tau_1,\tau_2,\tau_3\}$, donating a set of three distinctive textures: `Future', `Laser' and `Water'. For training set $S_{train}$, without loss of generality, we randomly chose 1 track from each of 161 genres via FMA dataset per iteration, and used the parts from 10s to 20s to generate content spectra. For testing set $S_{test}$ which was later utilized to evaluate our system, we selected a collection of five 10-second musical pieces.

\subsection{Metrics}
\subsubsection{Output Quality.}

We invited two human converters: $E$ was an engineer who is an expert in editing music while $A$ was an amateur enthusiast with three years' experience. They were asked to do the same task as ours: transferring music samples in $S_{test}$ to match texture samples in $S_{texture}$. For each task in $S_{test} \times S_{texture}$, we defined the output set ${S_{out}}_i$ as $\{E_i,A_i,M_i\}$, i.e., the output produced by $E$, $A$ and our network.

We considered using automatic score to compare our system to humans. However, it manifested that the machine evaluation could only measure one aspect of the transformation, and its effectiveness was upper bounded by the algorithm. As a result, we employed human judgment to assess the output quality of our system and human converters from three different dimensions: (1) texture conformity (2) content conformity (3) naturalness. 

To evaluate from both conformities, we collected Mean Opinion Score (MOS) from subjects using the CrowdMOS Toolkit\cite{crowdmos}. Listeners were asked to rate how well did music in ${S_{out}}_i$ match the corresponding sample in $t_{test}$ and $t_{testure}$ respectively. Specially, inspired by MUltiple Stimuli with Hidden Reference and Anchor (MUSHRA), we added the corresponding samples from $t_{test}$ and $t_{testure}$ as hidden reference. So as to better control the accuracy of our crowd-souring tests, apart from existing restrictions for MOS by ITU-T, answers which scored lower than 4 for hidden reference would also be automatically rejected.

As a crux of texture transfer tasks, naturalness of music produced by humans and our system was also marked. Since it was hard to score this property qualitatively, a Turing-test-like experiment was carried out. For each ${S_{out}}_i$, subjects were required to pick out the ``most natural (least awkward)'' sample. 

\subsubsection{Time-space Overhead.}
The computational performance of our system was evaluated, since it's one of the major determinants of user experience. We measured the average real execution time and maximal memory use with the production environment described above.

\subsection{Result and Analysis}\label{experiment results}

\begin{table*}[ht]
\caption{MOS scores (mean $\pm$ SD) for the conformity of content and texture, which are donated as $\Theta_c$ and $\Theta_t$ representatively.}\label{tab:conformity}
\centering
\noindent\begin{tabularx}{\textwidth}{XXXXXXX}
\toprule
\multirow{2}{*}{Converter} & \multicolumn{2}{c}{$\to \tau_1$} & \multicolumn{2}{c}{$\to \tau_2$} & \multicolumn{2}{c}{$\to  \tau_3$} \\ \cmidrule(lr){2-3}  \cmidrule(lr){4-5}  \cmidrule(lr){6-7} 
 & \multicolumn{1}{c}{$\Theta_c$} & \multicolumn{1}{c}{$\Theta_t$} & \multicolumn{1}{c}{$\Theta_c$} & \multicolumn{1}{c}{$\Theta_t$} & \multicolumn{1}{c}{$\Theta_c$} & \multicolumn{1}{c}{$\Theta_t$} \\ \midrule
E & $3.65 \pm 1.01$ & $3.62 \pm 0.83$ & $3.77 \pm 0.92$ & $3.71 \pm 1.02$ & $3.91 \pm 0.77$ & $3.59 \pm 0.87$ \\ \midrule
A & $3.19 \pm 1.26$ & $2.94 \pm 1.00$ & $3.10 \pm 1.05$ & $3.35 \pm 0.89$ & $3.18 \pm 1.15$ & $3.27 \pm 1.03$ \\ \midrule
Our & $2.97 \pm 1.17$ & $2.86 \pm 1.12$ & $2.96 \pm 1.13$ & $3.08 \pm 1.03$ & $3.22 \pm 1.12$ & $3.18 \pm 0.87$ \\ \bottomrule
\end{tabularx}
\end{table*}

\subsubsection{Output Quality.}
Results shown in Tab.~\ref{tab:conformity} indicate that, although scores for our output music are considerably lower than the ones scored for $E$ in conformity of both content and texture, they are close to the results of $A$. Specially, when transferring to $\tau_3$ (the texture of `Water'), our network even outperforms $A$ in preserving content information.

\begin{figure*}[htb]
\begin{center}
 \centerline{
 \includegraphics[width=\textwidth]{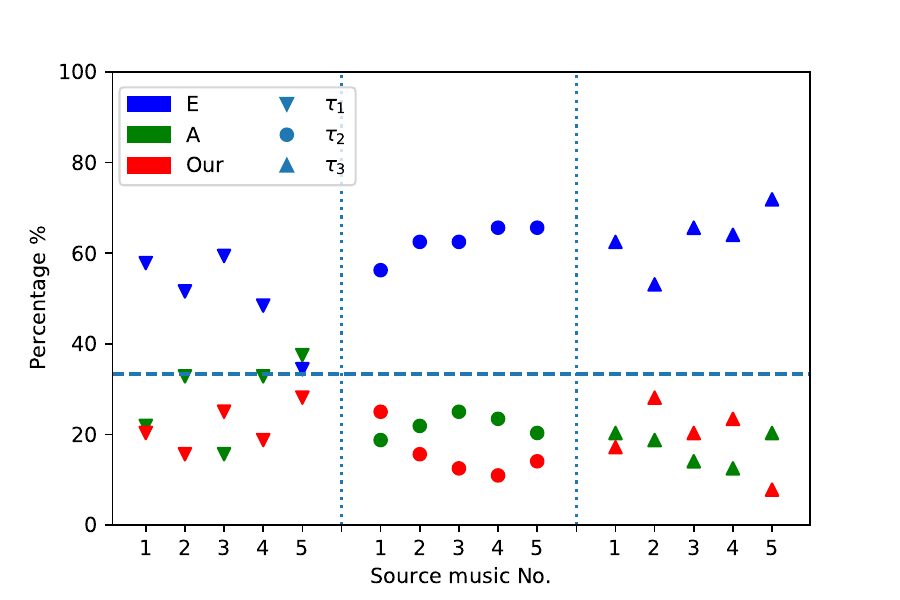}}
 \caption{The percentage of having the best naturalness for all tasks. The horizontal dashed line donates 33.3\% (random selection).}
 \label{fig:natrualness}
 \end{center}
\end{figure*}

Fig.~\ref{fig:natrualness} plots the results for naturalness test, which reveals that although there exists evident disparity between $E$ and the proposed system, there isn't much distinctness between the level of $A$ and ours.

\subsubsection{Time-space Overhead.}
During our experiment, the average runtime per transfer task is 30.84 seconds, and the memory use peak is 213 MB. The results validate that the overall computational performance meets the demand of the real-world application.

\begin{figure*}[ht]
\begin{center}
\subfloat[Pink noise (content)\label{fig:noise}]{\includegraphics[width=0.32\textwidth]{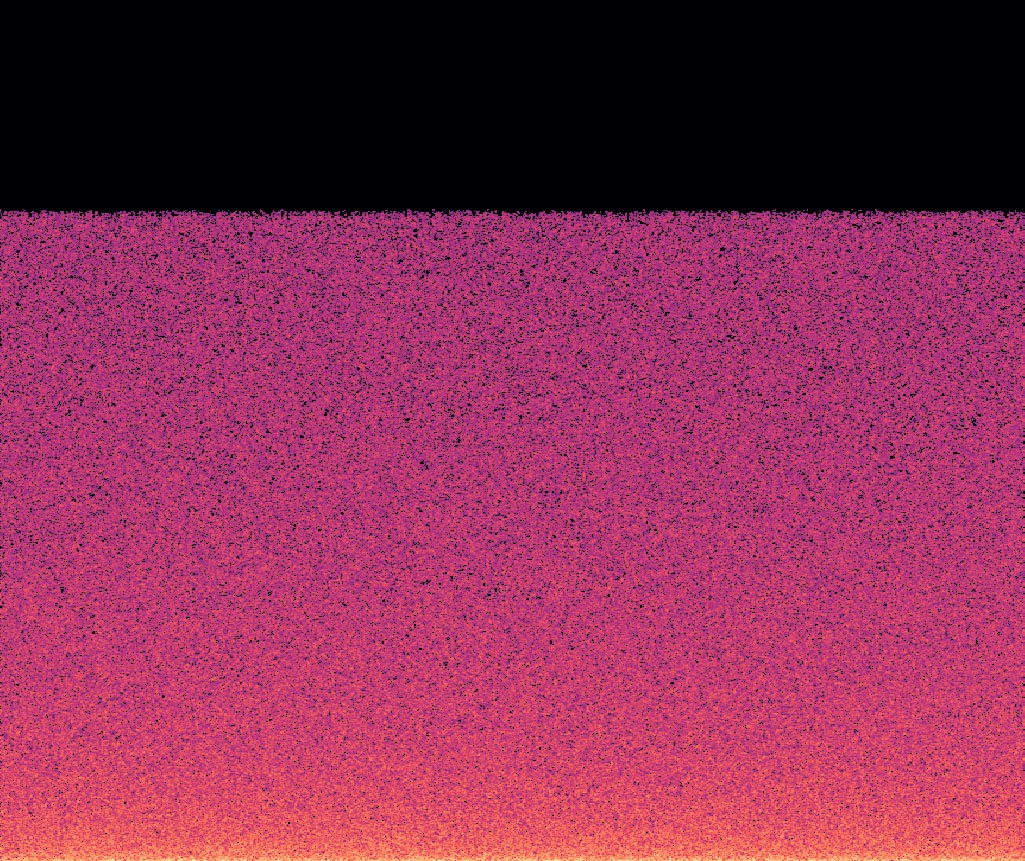}} \
\subfloat[Water (texture)\label{fig:water}]{\includegraphics[width=0.32\textwidth]{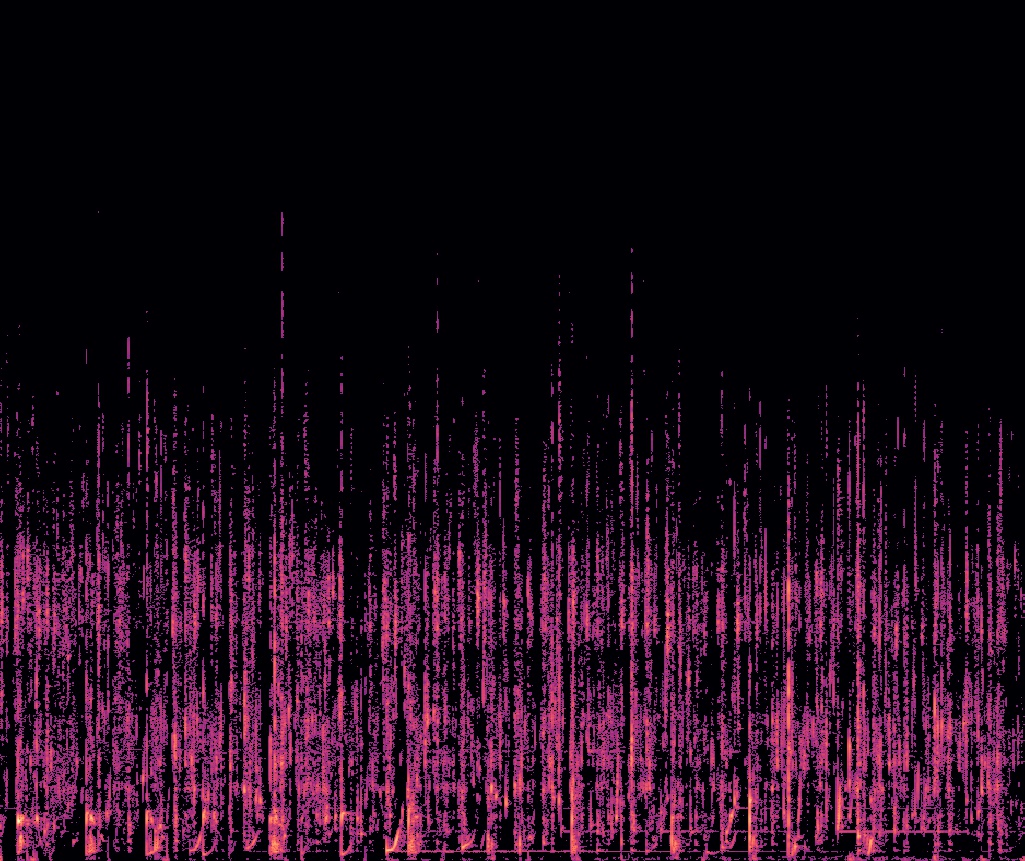}}\ 
\subfloat[Output result]{\includegraphics[width=0.32\textwidth]{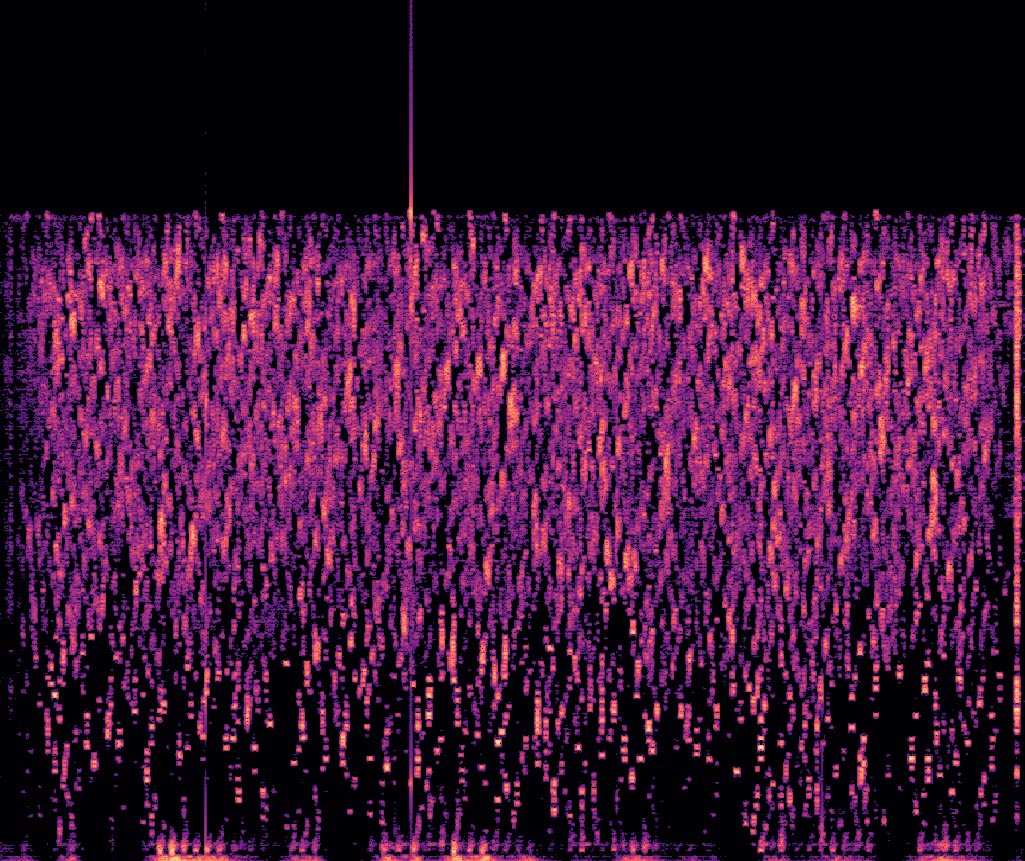}}  \\ 
 \caption{Spectral images used in texture synthesis evaluation. As shown in (c), the output audio of our method is fairly clear, i.e., most of the noisy `content' from (a) is gone. Moreover, it shares much texture features with (b).}  
 \end{center}
\end{figure*}\label{fig:ts}

\subsection{Byproduct: Audio Texture Synthesis}\label{ts}
The task of audio texture synthesis is to extract standalone texture feature from target audio, which is useful in sound restoration and audio classification. It's an interesting byproduct of our project, as it can be regarded as a special case of texture transfer when ridding the impacts of content audio (i.e., reduce $L_{content}$ to be 0). We generate pink noise pieces using W3C's Webaduio API\footnote{\url{https://www.w3.org/TR/webaudio/}} as content set $S_n$, select $\tau_3$ as target texture, and validate our system's effect of texture synthesis. Ideally, the influence from $S_n$ should be totally ruled out, while the repeated pattern from $\tau_3$ should appear. Qualitative results are shown in Fig.~\ref{fig:ts}, which revel our model's potential in audio texture synthesis.

\section{Conclusion and Future Work}
In this paper, we propose an end-to-end music texture transfer system.
To extract texture features, we first put forward a new reconstructive spectral representation on time-frequency.
Then, based on convolution operations, our network transfers the texture of music through processing its spectrum.
Finally, we rebuild the music pieces by utilizing proposed reconstructor.
Experimental results show that for texture transfer tasks, apart from the advantage of high performance, our deployed demo is on par with its amateur human counterparts in output quality. 

Our future work includes ameliorating of network structure, promoting training on other datasets and further utilizing our system for audio texture synthesis.

%
%
%
\bibliographystyle{splncs04}
\bibliography{cite}

\begin{thebibliography}{10}
\providecommand{\url}[1]{\texttt{#1}}
\providecommand{\urlprefix}{URL }
\providecommand{\doi}[1]{https://doi.org/#1}

\bibitem{barry2018style}
Barry, S., Kim, Y.: “style” transfer for musical audio using multiple
  time-frequency representations (2018),
  \url{https://openreview.net/forum?id=BybQ7zWCb}

\bibitem{stargan}
Choi, Y., Choi, M., Kim, M., Ha, J.W., Kim, S., Choo, J.: Stargan: Unified
  generative adversarial networks for multi-domain image-to-image translation.
  In: Proc. of CVPR (2018)

\bibitem{fma_dataset}
Defferrard, M., Benzi, K., Vandergheynst, P., Bresson, X.: Fma: A dataset for
  music analysis. In: Proc. of ISMIR (2017)

\bibitem{nlptf}
Fu, Z., Tan, X., Peng, N., Zhao, D., Yan, R.: Style transfer in text:
  Exploration and evaluation. In: Proc. of AAAI (2018)

\bibitem{Gatys}
Gatys, L.A., Ecker, A.S., Bethge, M.: Image style transfer using convolutional
  neural networks. In: Proc. of CVPR (2016)

\bibitem{goldstein2014sensation}
Goldstein, E.: Sensation and perception. Wadsworth, Cengage Learning (2014)

\bibitem{GLA}
Griffin, D., Lim, J.: Signal estimation from modified short-time fourier
  transform. IEEE Transactions on Acoustics, Speech, and Signal Processing
  \textbf{32}(2) (1984)

\bibitem{deepbach}
Hadjeres, G., Pachet, F., Nielsen, F.: Deepbach: a steerable model for bach
  chorales generation. In: Proc. of ICML (2017)

\bibitem{sentiment}
Hu, Z., Yang, Z., Liang, X., Salakhutdinov, R., Xing, E.P.: Toward controlled
  generation of text. In: Proc. of ICML (2017)

\bibitem{p_style}
Jhamtani, H., Gangal, V., Hovy, E., Nyberg, E.: Shakespearizing modern language
  using copy-enriched sequence-to-sequence models. In: Proc. of the Workshop on
  Stylistic Variation (2017)

\bibitem{Johnson2016Perceptual}
Johnson, J., Alahi, A., Fei-Fei, L.: Perceptual losses for real-time style
  transfer and super-resolution. In: Proc. of ECCV (2016)

\bibitem{Malik}
Malik, I., Ek, C.H.: Neural translation of musical style. In: Proc. of the NIPS
  Workshop on ML4Audio (2017)

\bibitem{MCDERMOTT2011926}
McDermott, J., Simoncelli, E.: Sound texture perception via statistics of the
  auditory periphery: Evidence from sound synthesis. Neuron  \textbf{71}(5)
  (2011)

\bibitem{Mital}
Mital, P.K.: Time domain neural audio style transfer. In: Proc. of the NIPS
  Workshop on ML4Audio (2017)

\bibitem{fair}
Mor, N., Wolf, L., Polyak, A., Taigman, Y.: A universal music translation
  network. CoRR  \textbf{abs/1805.07848} (2018)

\bibitem{Oord2016WaveNetAG}
van~den Oord, A., Dieleman, S., Zen, H., Simonyan, K., Vinyals, O., Graves, A.,
  Kalchbrenner, N., Senior, A.W., Kavukcuoglu, K.: Wavenet: A generative model
  for raw audio. In: SSW (2016)

\bibitem{crowdmos}
P.~Ribeiro, F., Florencio, D., Zhang, C., Seltzer, M.: Crowdmos: An approach
  for crowdsourcing mean opinion score studies. In: Proc. of ICASSP (2011)

\bibitem{VGG19}
Simonyan, K., Zisserman, A.: Very deep convolutional networks for large-scale
  image recognition. CoRR  \textbf{abs/1409.1556} (2014)

\bibitem{Ulyanov2016InstanceNT}
Ulyanov, D., Vedaldi, A., Lempitsky, V.S.: Instance normalization: The missing
  ingredient for fast stylization. CoRR  \textbf{abs/1607.08022} (2016)

\bibitem{Wyse2017}
Wyse, L.: Audio spectrogram representations for processing with convolutional
  neural networks. In: Proc. of DLM2017 joint with IJCNN (2017)

\bibitem{texture_images}
Yu, G., Slotine, J.J.E.: Audio classification from time-frequency texture. In:
  Proc. of ICASSP (2009)

\bibitem{CycleGAN2017}
Zhu, J.Y., Park, T., Isola, P., Efros, A.A.: Unpaired image-to-image
  translation using cycle-consistent adversarial networks. In: Proc. of ICCV
  (2017)

\end{thebibliography}

\end{document}